\documentclass{revtex4}

\usepackage{graphicx}
\usepackage{dcolumn}
\usepackage{bm}
\usepackage{amsmath}
\usepackage{amssymb}
\usepackage{amsfonts}
\usepackage{color}
\usepackage{mathtools}
\usepackage{nicefrac}
\usepackage{subfigure}

\newcommand{\period}{\hspace{0.005\linewidth} . \hspace{0.02\linewidth} }
\newcommand{\coma}{\hspace{0.02\linewidth} , \hspace{0.02\linewidth} }
\newcommand{\with}{\hspace{0.02\linewidth} \text{ with } \hspace{0.02\linewidth} }
\newcommand{\andd}{\hspace{0.02\linewidth} \text{ and } \hspace{0.02\linewidth} }
\newcommand{\for}{\hspace{0.02\linewidth} \text{ for } \hspace{0.005\linewidth} }

\newcommand{\bs}[1]{\boldsymbol{#1}}
\newcommand{\rey}[1]{\overline{#1}}
\newcommand{\integ}[1]{\big\langle #1 \big\rangle}

\newcommand{\ta}{t}
\newcommand{\ua}{u}

\newcommand{\uaf}{\widehat{\ua}}

\newcommand{\tf}[1]{\widehat{#1}}
\newcommand{\pp}{\mathcal{P}}
\newcommand{\proj}{P}
\newcommand{\nk}{\hat{k}}

\newcommand{\bnk}{\bs{\hat{\K}}}

\newcommand{\x}{\bs{x}}
\newcommand{\K}{\bs{k}}
\newcommand{\p}{\bs{p}}
\newcommand{\q}{\bs{q}}

\newcommand{\ud}{\mathrm{d}}
\newcommand{\cjgt}{*}

\def\figpath{./figures/}
\def\figpath2{./}

\date{\today}

\begin{document}

\title{
 Permanence of large eddies in variable-density homogeneous turbulence
}

\author{O. Soulard}
\author{J. Griffond}
\author{B.-J. Gr\'ea}
\author{G. Viciconte}
\affiliation{CEA, DAM, DIF, F-91297 Arpajon, France.}

\begin{abstract}
The principle of permanence of large eddies is one of the central pillars onto which our understanding of decaying homogeneous turbulence is built. The validity conditions of this principle have been thoroughly discussed for constant density flows, but not for variable-density ones. 
In this work, we show that density non-uniformities modify the remote action of the pressure field. It results into distant velocity correlations being submitted to a stronger non-linear transfer of energy.  A simple example is proposed to illustrate this property and a spectral analysis of non-linear transfer terms is undertook to further characterize it. From there, we derive that large eddies in variable density flows remain permanent for a smaller set of initial conditions than when density is constant. Permanence strictly applies to initial spectra having an infrared exponent smaller than 2 instead of 4.
Implicit large-eddy simulations are performed to verify the main predictions of this work.
\end{abstract}

\maketitle

\section{Introduction.~}
Large eddies -- eddies much larger than the integral scale of turbulence -- play a central role in the  decay of constant density homogeneous turbulence \cite{loitsyanskii39,kolmogorov41b,landau54,saffman67,lesieur00,davidson04,lesieur08,llor11,mons14}.
Under certain conditions, detailed in the next paragraph, large eddies evolve on a time scale much longer than the one governing the decrease of kinetic energy or the growth of the integral scale. 
As a result, their initial state is preserved during the whole flow evolution: they are said to be permanent.
When this permanence is verified, the flow is constrained by its large-scale initial condition and its decay rate can be expressed as a function of some large-scale initial parameter, such as, for instance, the initial infrared exponent.
The latter corresponds to the power law exponent of the initial spectrum at small wave numbers and is denoted by $s_0$.

Whether large eddies are permanent or not depends on the long range energy transfers induced by non-linear terms. 
In a constant density flow, these terms involve  quadratic products between the  components of the velocity field, and their non-local propagation by the pressure field. 
When $s_0 \ge 2$, these non-linear terms lead to a ``backscattering'' transfer of energy: large eddies receive energy from interactions involving smaller eddies, mostly those with a size on the order of the integral scale.
Several models \cite{proudman54,lesieur08,llor11} predict that this backscattering transfer has an infrared exponent of $4$ when expressed in spectral space.
Thus, for initial spectra satisfying $s_0<4$, non-linear processes have a vanishingly small effect and the infrared spectrum is invariant, i.e. large eddies are permanent.
By contrast, spectra with $s_0 > 4$ are not invariant: they transition to a spectrum with an infrared exponent of $4$, equal to that of the backscattering term.

When density is not constant, a major difference arises: a non-linearity that does not exist in the constant density case must be accounted for.
Along with the usual quadratic term involving the velocity field, a non-linear product between the inverse density and the pressure gradient is also active.
This additional non-linearity affects the way the pressure field transmits information over very long distances: the Poisson equation for pressure is indeed modified by its presence.
Therefore, the long range correlations of the velocity field are possibly impacted by this additional non-linear term, along with the conditions under which large eddies are permanent.
The purpose of this work is to examine this issue.

\vspace{0ex}
\section{A simple example.~}
The pressure field is one of the primary factors involved in the evolution of the statistical properties of large eddies. Indeed, because of its non-local nature, it is at the origin of long range correlations between distant points.
In this respect, a simple example allows to shed light on the remote action of the pressure field and to gain insight into the properties of large eddies in homogeneous turbulence.
This example was first introduced by  Batchelor \& Proudman \cite{batchelor56}.
The latter considered a ``blob'' of vorticity located close to the origin $\x=0$ of an infinite domain and analyzed the pressure and velocity fields far away from the origin. 
Here, we add an extra-element compared to \cite{batchelor56}: within the domain $\mathcal{D}$ where vorticity is non-zero, we assume that the density field $\rho$ is non-uniform, while outside of $\mathcal{D}$, it is constant. 

Whether the density is variable or not, 
the velocity $\bs{u}$ is incompressible and is related to the vorticity field $\bs{\omega}$ by the expression:
$$
4 \pi {u}_i(\x) = \epsilon_{ijk} \partial_k \int \omega_j(\x') \frac{\ud \x'}{\left|\x - \x' \right|}
\period
$$
Far from the blob, this expression can be simplified by Taylor-expanding $\left|\x - \x' \right|$. Denoting the characteristic length of the domain $\mathcal{D}$ by $\ell_\mathcal{D}$, one obtains that \cite{davidson04}:
\begin{align} \label{eq:u_farfield}
\text{for }  |\x| \gg \ell_\mathcal{D} \coma 4 \pi {u}_i(\x) =  L_j \partial^2_{ij}(|\x|^{-1}) 
\\ \nonumber
+ \frac{1}{2} \epsilon_{ijk} \partial^3_{kpq}(|\x|^{-1}) \int x'_p x'_q \omega_j(\x') \ud \x'
\coma
\end{align}
with $\bs{L}$ the linear impulse of the blob of vorticity:
$L_i = \frac{1}{2} \int \epsilon_{ijk} x_j \omega_k \ud \x$ \cite{davidson04}.
Thus, the far-field velocity scaling depends on whether the eddies within the vorticity region have a linear impulse or not:
$$
\text{if } \bs{L} \ne 0 \coma \bs{u}(\x) \sim |\x|^{-3} 
\andd 
\text{if } \bs{L} = 0 \coma \bs{u}(\x) \sim |\x|^{-4}
\period 
$$
Without density variations, it has been shown that the linear impulse is an invariant of an isolated blob of vorticity \cite{davidson04}. The pressure decays sufficiently fast at infinity so that it does not affect the evolution of $\bs{L}$.
In particular, if $\bs{L}$ is initially null, it remains so  and $\bs{u}(\x) \sim |\x|^{-4}$ at all times.
However, when density variations exist, the situation is different.
To show this, let us start from the incompressible Navier-Stokes equations:
\begin{align} \label{eq:blob_NS}
\partial_t u_i + u_j\partial_j u_i = - \tau \partial_i p + \nu \partial^2_{jj} u_i
\coma
\partial_j u_j = 0
\coma
\end{align}
with $p$ the pressure field, $\tau=1/\rho$ the specific volume and $\nu$ a constant viscosity.
The incompressibility constraint leads to the following Poisson equation for the pressure:
\begin{equation} \label{eq:blob_poisson}
\partial^2_{jj} \left(\tau p\right) - \partial_j\left( p \partial_j \tau\right)= -\partial^2_{ij}\left( u_i u_j\right) 
\period
\end{equation}
This equation is different from the one obtained in the constant density case.
Indeed, it is not a Laplacian which acts on the pressure because $\tau$ is not constant.
This difference is such that an expression for the pressure as a function of the velocity field alone cannot be found in general.
Still, an implicit solution can be expressed by inverting the Laplacian acting on $\tau p$. This leads to:
$$
4 \pi  \,\tau p(\x) = \int  \partial_j\big[ \partial_i(u_iu_j) - p \partial_j \tau \big](\x') \frac{\ud \x'}{\left|\x - \x' \right|}
\period
$$
By expanding $\left|\x - \x' \right|$, we obtain that, far from the blob:
\begin{align} \label{eq:blob_press}
& \text{for } |\x| \gg \ell_\mathcal{D} \coma 
4 \pi  \,\tau p(\x) = - \partial_{i}(|\x|^{-1}) \int p \partial_i \tau \;\ud \x 
  \\ \nonumber
& +
\partial^2_{ij}(|\x|^{-1}) \left( \int u_iu_j \ud \x- \frac{1}{2} \int p \left( x_i \partial_j\tau   + x_j \partial_i\tau \right)\; \ud \x \right)
\period
\end{align}
The integrals involving $p \partial_i \tau$ converge because we assumed that $\tau$ is constant outside the domain $\mathcal{D}$, so that $\partial_i \tau = 0$ for $\x \notin \mathcal{D}$.
In the constant density case, one simply has $4 \pi  \,\tau p(\x) = \partial^2_{ij}(|\x|^{-1}) \, \int u_iu_j \ud \x$.
Therefore, the scaling of the pressure field is not the same in the constant and variable density cases. 
Another distinction must also be made. The size of the blob $\ell_\mathcal{D}$ is not any more the sole length characterizing the problem.
The comparison of the prefactors of the two terms in Eq. \eqref{eq:blob_press} leads to the definition of another scale $\ell_P$ defined as:
$$
\ell_P = \frac{\int u_iu_i \ud \x}  
{|\int p |\bs{\partial}_{\bs{x}} \tau| \;\ud \x|}
\period
$$
With this definition, we can write that far from the vorticity blob, $p$ verifies:
\begin{align}
\text{for } |\x| \gg \ell_\mathcal{D}, \ell_P  \coma
p(\x) \sim |\x|^{-2}
\period
\end{align}
This result must be contrasted with the constant density scaling: $p(\x) \sim |\x|^{-3}$.
Density variations within the vorticity blob induce a pressure field decaying more slowly than in the constant density case.
This difference has strong implications for the conservation of the linear impulse.
Indeed, injecting the far-field expansion of $\bs{u}$ given by Eq. \eqref{eq:u_farfield} into its evolution equation \eqref{eq:blob_NS}, and using the scaling obtained for the pressure, we deduce that:
\begin{align} \label{eq:lin_imp}
   \partial_t L_i = \int p \partial_i \tau \;\ud \x 
\period
\end{align}
Note that this result can also be derived by injecting the definition of $\bs{L}$ into the vorticity equation, but without gaining any insight into the pressure scaling. 
If we except special cases, like barotropic flows, Eq. \eqref{eq:lin_imp} implies that, in general, $\partial_t \bs{L} \ne 0$. 
In particular, when $\bs{L}$ is initially null, it does not stay so:
 linear impulse is created by the correlation between pressure and density gradient  within the vorticity blob.
As a result, when density is variable, one has:
$$
\bs{L} \ne 0 \andd \bs{u}(\x) \sim |\x|^{-3} \for |\x| \gg \ell_\mathcal{D}, \ell_P
\period 
$$
The previous results are valid for the far-away pressure and velocity fields, located at distances from the origin much larger than both $\ell_\mathcal{D}$ and $\ell_P$.
In this respect,  an important aspect was left out of the discussion: the possibility of the existence of an intermediate large-scale range between $\ell_\mathcal{D}$ and $\ell_P$.
Such a range can exist provided $\ell_P \gg \ell_\mathcal{D}$.
In this intermediate range, $p$ obeys the following scaling according to Eq. \eqref{eq:blob_press}:
\begin{align*}
& \text{for } \ell_\mathcal{D} \ll  |\x| \ll \ell_P  \coma p(\x) \sim |\x|^{-3}
\period
\end{align*}
Thus, for $\ell_P\gg\ell_\mathcal{D}$, we distinguish a first range of scales where the pressure field obeys its usual constant density scaling. In this range, the velocity field  scales as $|\x|^{-4}$ if $\bs{L}$ is initially null.
At larger distances, for $|\x| \gg \ell_P$, the asymptotic results derived in the first part of this section apply: the velocity field evolves as $|\x|^{-3}$ even if $\bs{L}$ is initially null.
Of course, when $\ell_P \gtrsim \ell_\mathcal{D}$, only the range predicted in the first part of this section appears.

The order of magnitude of $\ell_\mathcal{D}/\ell_P$ is hard to assess in the general case. To understand how this ratio might vary, we consider two examples. 
In the first one, we assume that the characteristic length of density, velocity  and their gradients are on the order of $\ell_\mathcal{D}$. 
This hypothesis is relevant to the early times of a turbulent mixing flow, when turbulence has not yet set in and small scales are absent. 
Then, the following order of magnitude should be verified:
$$
\ell_\mathcal{D}/\ell_P \sim {\sqrt{\integ{\tau'^2}}_\mathcal{D}}/{\integ{\tau}_\mathcal{D}}
\period
$$
where $\integ{\cdot}_\mathcal{D}$ refers to the average over the blob and where the prime refers to the fluctuation with respect to the average $\integ{\cdot}_\mathcal{D}$.
Therefore, for small density variations, i.e. for $\nicefrac{\sqrt{\integ{\tau'^2}}_\mathcal{D}}{\integ{\tau}_\mathcal{D}} \ll 1$, one should have $\ell_\mathcal{D}/\ell_P \ll 1$ and two large-scale ranges should exist. 
The other example we consider is when turbulence has set in within the blob and when the Reynolds number is very high. Then, with $p$ a large scale quantity and $\partial_i \tau$ a small scale one, we expect a decorrelation between the two quantities, linked to the Reynolds number $Re$ so that $\integ{p \partial_i \tau}_\mathcal{D}$ becomes negligible. Hence, we expect that;
$$
\text{For } Re \gg 1 \coma \ell_\mathcal{D}/\ell_P \ll 1
\period
$$
Therefore, we anticipate that the effects of density variations on the pressure scaling will decrease as the Reynolds number increases.

\vspace{0ex}
\section{Application to homogeneous turbulence.~}
The results derived so far for an isolated eddy can be transposed to homogeneous turbulence.
As shown in \cite{saffman67,llor11}, a collection of independent eddies having a linear impulse leads to a Saffman turbulent spectrum, scaling as $k^{2}$, with $k$ the wave number. By contrast, a collection of eddies without linear impulse gives rise to a Batchelor spectrum, scaling as $k^4$.
We showed that, with density variations, the linear impulse of an eddy is not zero and is time dependent. Therefore, if one considers a homogeneous sea of independent eddies with density fluctuations, one should eventually observe a Saffman $k^2$ spectrum even if initially it verifies Batchelor $k^4$ scaling. Besides, if the initial spectrum has a Saffman scaling then its large scales are not permanent, even though the scaling exponent is preserved.
By contrast, without density variations, a Batchelor spectrum would remain so and a Saffman spectrum would be permanent.
To sum up, the permanence of large eddies would only be reached for $s_0 <2$ for a variable-density flow, while it is reached for $s_0<4$ for a constant density flow.

To confirm these expectations, we propose to analyze, in spectral space,  a homogeneous decaying variable-density turbulent flow. 
We introduce $\bs{\uaf}(\K,t)$  the Fourier transform of the velocity field at a given wave vector $\K$ and at time $t$.
Applying the Fourier transform to  Eq. \eqref{eq:blob_NS} for the velocity and to Eq. \eqref{eq:blob_poisson} for the pressure, we deduce that:
\begin{align}  \label{eq:NS_tf}
\partial_t \tf{u}_i  = 
- \imath \frac{k}{2}  \pp_{ijk}(\bnk) \tf{\ua_j\ua_k}  
+ P_{ij}(\bnk) \tf{p \partial_j \tau}
 - \nu k^2 \tf{u}_i
\coma
\end{align}
with $\proj_{ij}$ and $\pp_{ijk}$ projectors on incompressible fields:
$$
\proj_{ij}(\bnk) = \delta_{ij} - {\nk}_i {\nk}_j
\coma
\pp_{ijk}(\bnk) = \nk_j \proj_{ik}(\bnk) + \nk_k \proj_{ij}(\bnk)
\period
$$
with $\bnk=\K / k$ the direction of the wave vector $\K$.
The first and last terms on the right-hand side are obtained whether $\tau$ is uniform or not.
The second one appears only when density is variable, and its role is similar to the one played by the integral $\int\! p \partial_i \tau \ud \x$  in the vorticity blob case.

Our focus is on the properties of the kinetic energy spectral density: 
$
E(\K,t) \delta\left(\K-\K'\right)
= \frac{1}{2} \rey{\tf{\ua_i}(\K,\ta) {\tf{\ua_i}^{\cjgt}(\K',\ta)}} 
\period
$
Its evolution equation can be formally written as:
\begin{align}
& \partial_t E(\K,t) =  T^{(0)}(\K,t) + \int_0^t T(\K,t,s)  \ud s\coma
\\
\nonumber
\with &  
T^{(0)}(\K,t)\delta\left(\K-\K'\right)
= \Re\left( \rey{\tf{u}_i^{(0)}\hspace{-0.5ex}(\K) \partial_t\tf{u}^\cjgt_i(\K',t)   } \right)
\\
\nonumber
\andd &
T(\K,t,s) \delta\left(\K-\K'\right)
= \Re\left( \rey{\partial_t\tf{u}_i(\K,t)\partial_t\tf{u}^\cjgt_i(\K',s) } \right)
\coma
\end{align}
with $\tf{u}_i^{(0)}$ the value of $\tf{u}_i$ at $t=0$ and $\Re$ the real part. 
Neglecting viscous effects, the second component of the transfer term can be expressed as:
\begin{align} \label{eq:Tnl}
& \rey{\partial_t\tf{u}_i(\K,t)\partial_t\tf{u}^\cjgt_i(\K',s)} =
\\ \nonumber
&
k^2 \; \nk_i \nk_k P_{jl}(\bnk) \; \rey{\tf{\ua_i\ua_j}(\K,t)\tf{\ua_k\ua_l}^\cjgt(\K',s) } 
\\ \nonumber
& - \frac{\imath k}{2}\mathcal{P}_{ijk} \!\!\left( \rey{ \tf{\ua_j\ua_k}(\K,t)  \tf{p \partial_i \tau}^\cjgt(\K',s)} \!-\! \rey{ \tf{\ua_j\ua_k}^\cjgt(\K',s)  \tf{p \partial_i \tau}^\cjgt(\K,t)}  \right)
\\ \nonumber
& + 
P_{ij}(\bnk) \; \rey{ \tf{p \partial_j \tau}(\K,t)\tf{p \partial_i \tau}^\cjgt(\K',s) }
.
\end{align}
In order to model these non-linear terms at large scales, 
we assume that the spectra of all the fluctuating quantities peak at a similar wave number $k_e(t)$.
Besides, we assume that the largest contribution to the kinetic energy and other fluctuation variances comes from a range of wavenumbers close to or larger than $k_e(t)$, while smaller wavenumbers only provide a marginal contribution. 
This energy  containing range is denoted by $k \gtrsim k_e(t)$.  The large scale range, the one we are interested in, is defined as $k \ll k_e(t)$.
Note that, as time increases, the integral scale of turbulence increases, i.e. $k_e(t)$  decreases. Thus, a wave number belonging to the large scale range at time $t$ also belongs to the large scale range at time $s<t$.

Using this assumption, we can now simplify the expressions of the fourth order correlations defining $T$.
For instance, we can write that, for $k \ll k_e(t)$:
\begin{align*}
& \rey{\tf{\ua_i\ua_j}(\K,t)\tf{\ua_k\ua_l}^\cjgt(\K',s) } =
\\ 
&  \hspace{3ex} \iint S_{ijkl}(\p, \K\!-\!\p,\q,\K'-\q;t,s) \ud \p \ud \q \; \delta(\K-\K')
\\ 
&\approx 
\iint_{p \gtrsim k_e(t) , q \gtrsim k_e(s)} \hspace{-12ex} S_{ijkl}(\p, \K\!-\!\p,\q,\K'-\q;t,s) \ud \p \ud \q \; \delta(\K-\K')
\\ 
& \approx 
\iint_{p \gtrsim k_e(t) , q \gtrsim k_e(s)} \hspace{-12ex} S_{ijkl}(\p, \!-\!\p,\q,-\q;t,s) \ud \p \ud \q \; \delta(\K-\K')
\coma
\end{align*}
with $S_{ijkl}(\bs{a}, \bs{b},\bs{c},\bs{d}; t,s) \delta(\bs{a}+\bs{b} \!-\!\bs{c}\!-\!\bs{d})\!=\!\rey{\tf{\ua}_i(\bs{a},t)\tf{\ua}_j(\bs{b},t)\tf{\ua}_k^\cjgt(\bs{c},s)\tf{\ua}_l^\cjgt(\bs{d},s)}$.
The first equality is the definition of the convolution product. The first approximation is a direct expression of our main assumption and the second one is a Taylor expansion in the limit $k\ll k_e$.
The end result here is that $\rey{\tf{\ua_i\ua_j}(\K,t)\tf{\ua_k\ua_l}^\cjgt(\K',s) }$ depends only on $t$ and $s$ but not on the wave vector $\K$.
Similar conclusions apply to the other fourth order correlations.
We  then deduce that:
\begin{align*}
T(\K,t,s) = k^2 \mathcal{T}^{(1)}(\bnk,t,s) + k \mathcal{T}^{(2)}(\bnk,t,s) + \mathcal{T}^{(3)}(\bnk,t,s)
\coma
\end{align*}
where $\mathcal{T}^{(1)}$, $\mathcal{T}^{(2)}$ and $\mathcal{T}^{(3)}$ correspond respectively to the first, second and third terms of the right-hand side of Eq. \eqref{eq:Tnl}.

As a  last hypothesis, we assume that the initial value $\tf{u}^{(0)}$ is uncorrelated with the time derivative of $\tf{u}$. Hence, we set $T^{(0)}=0$.
Finally, we obtain the following modeled evolution for $E$ at large scales $k\ll k_e(t)$:
\begin{align}
\partial_t E(\K,t) = & k^2 \int_0^t \mathcal{T}^{(1)}(\bnk,t,s)\ud s + k \int_0^t \mathcal{T}^{(2)}(\bnk,t,s) \ud s
\nonumber \\ &
+ \int_0^t \mathcal{T}^{(3)}(\bnk,t,s) \ud s
\period
\end{align}
In a constant density flow, only the first term is present: non-linear interactions have a spectral density scaling as $k^2$ or equivalently a modulus spectrum scaling as $k^4$.
When density is variable, additional non-linear terms, stemming from the correlations between pressure and density gradients, arise.
The leading order term $\mathcal{T}^{(3)}$ has a spectral density scaling as $k^0$ or equivalently a modulus spectrum scaling as $k^2$. 

This difference mirrors the various pressure scalings derived in the vorticity blob case.
It has important consequences concerning the permanence of large eddies.
Indeed, initial spectra with $s_0 > 2$ are not invariant: they transition to a spectrum with an infrared exponent of $2$. The permanence of large eddies is only achieved for initial spectra satisfying $s_0<2$.

This conclusion must be mitigated by the following consideration. As in the vorticity blob case, the comparison of the orders of magnitude of $\mathcal{T}^{(i)}$ leads to the definition of an additional characteristic wavenumber:
$
k_p(t) = {|\rey{p |\bs{\partial}_{\bs{x}} \tau|}|}/
{\rey{u_iu_i} }\period
$
Then, if $k_p \ll k_e$, two large-scale ranges exist, verifying
\begin{align*}
& \text{for } k_p \ll k \ll k_e \coma && T(\K,t,s) \propto k^2 
\\
\andd &
\text{for } k \ll k_p \coma &&T(\K,t,s) \propto k^0
\period
\end{align*}
In particular, when $s_0<4$ and when $k_p \ll k_e$, one should observe a kinetic energy spectrum scaling as $k^4$ for $k_p \ll k \ll k_e$ and $k^2$ for $k\ll k_p$.
The conditions under which $k_p \ll k_e$ are similar to the ones discussed in the vorticity blob case. They are verified, among others, when the density contrast is small : $\nicefrac{\sqrt{\rey{\tau'^2}}}{\rey{\tau}} \ll 1$.
Since, this quantity decays with time, there is a possibility that spectra with neither a $k^4$ or $k^2$ scaling are observed.

\vspace{0ex}
\section{Validation.~}
\begin{figure}[!tb]
  \includegraphics[width=0.5\linewidth]{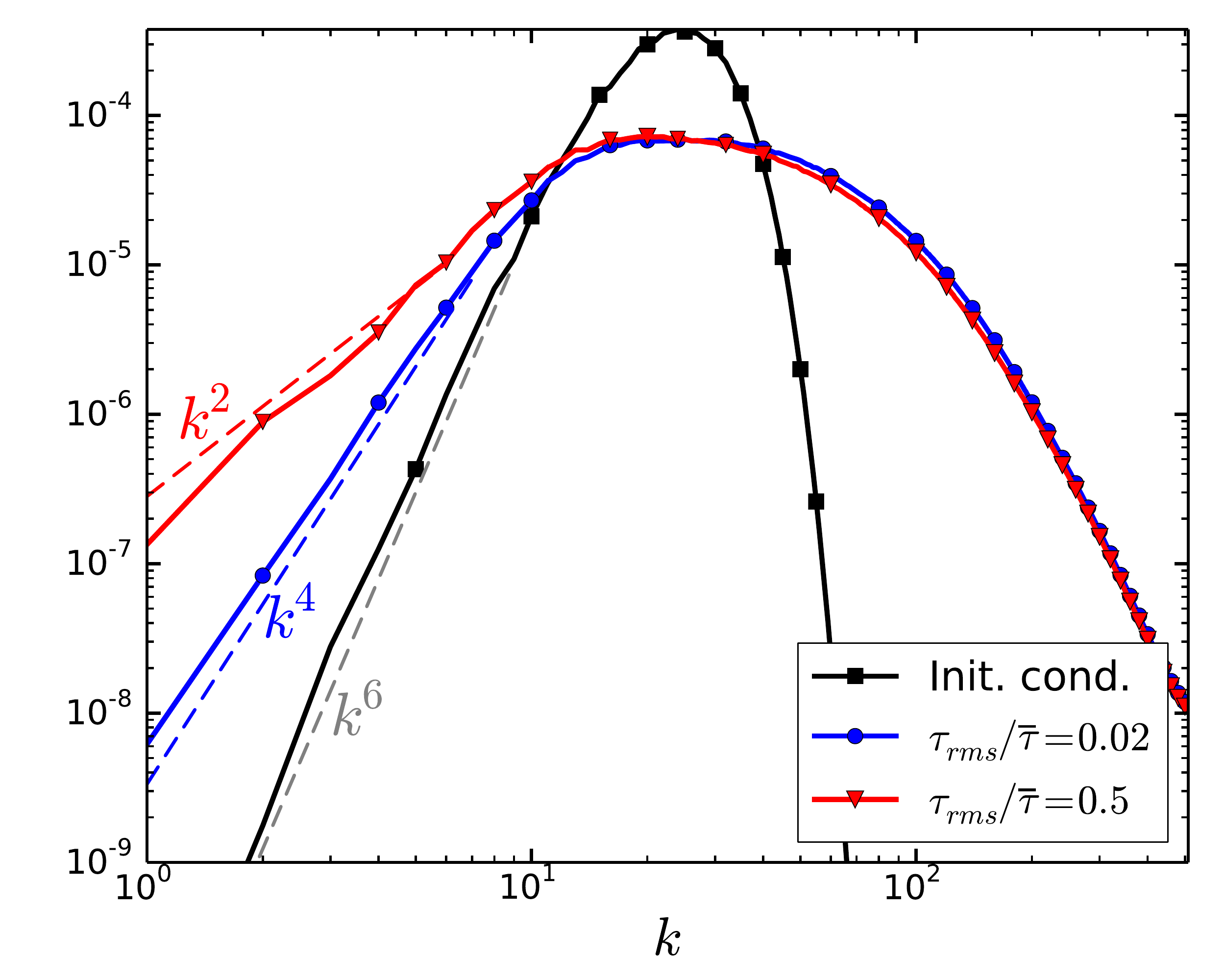}
\caption{\label{fig:Epp_9f}
Kinetic energy spectrum at $t=40 /(k_0 \left.\sqrt{\rey{u_iu_i}}\right|_0)$.
}
\end{figure}

\begin{figure}[!tb]
  \includegraphics[width=0.5\linewidth]{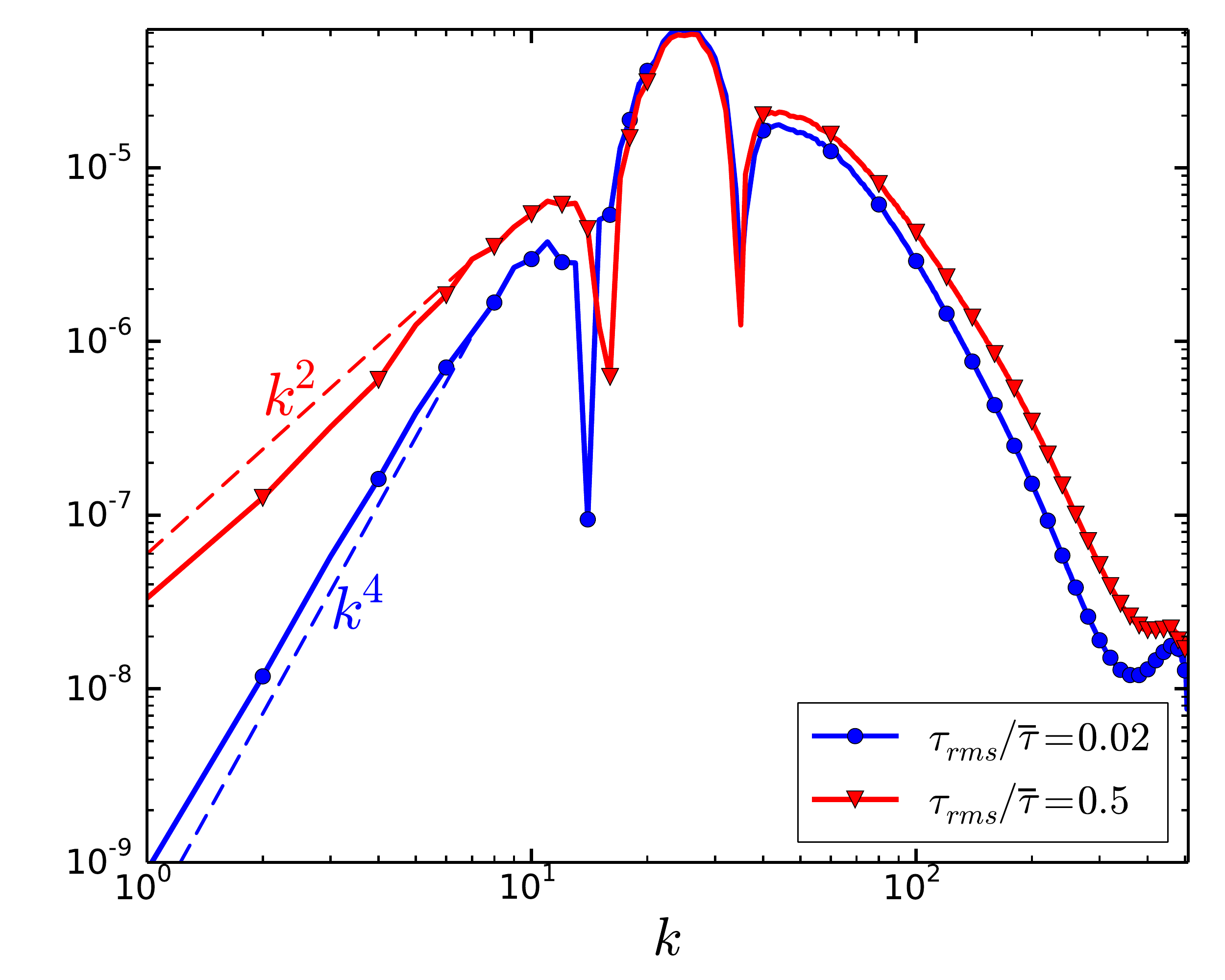}
\caption{\label{fig:NL_9f}
Absolute value of the non-linear transfer term at $t=10 /(k_0 \left.\sqrt{\rey{u_iu_i}}\right|_0)$. 
}
\end{figure}

To illustrate some of our predictions, we perform implicit large eddy simulations of homogeneous isotropic turbulence with the code \textsc{Triclade} \cite{shanmuganathan13}.
In a box of size $2\pi$, discretized by $1024^3$ cells, we impose an initial kinetic spectrum proportional to $(k/k_0)^6 \exp(-(k/k_0)^2)$, with $k_0=20$. The initial specific volume spectrum is uncorrelated and follows the same variation.
 Two simulations are done, one with  an initial value $\nicefrac{\sqrt{\tau'^2}}{\rey{\tau}}=0.5$ and the other with $\nicefrac{\sqrt{\tau'^2}}{\rey{\tau}}=0.02$.
It can be seen in Fig. \ref{fig:Epp_9f} that the spectrum transitions rapidly towards a $k^2$ spectrum when the density contrast is high and to a $k^4$ spectrum when this contrast is small. Correspondingly, the non-linear transfer term, displayed at an intermediate time in Fig. \ref{fig:NL_9f}, exhibits a $k^2$ infrared scaling for $\nicefrac{\sqrt{\tau'^2}}{\rey{\tau}}=0.5$ and a $k^4$ scaling for $\nicefrac{\sqrt{\tau'^2}}{\rey{\tau}}=0.02$.
These elements agree with our predictions.

\section{Conclusions}
In this work, we studied the influence of density variations on the large scales of homogeneous turbulence.
We showed that density variations modifiy the transfer of kinetic energy at large scales, by introducing an additional non-linear component scaling as $k^2$ for small $k$. As a result, the permanence of large eddies is only ensured for initial spectra having an infrared exponent $s_0 < 2$. For $s_0>2$, the outcome depends on the density contrast and on other parameters characterizing the initial state. 
Those predictions were verified by performing implicit large eddy simulations of homogeneous turbulence.


\end{document}